\documentclass[journal, twocolumn, 12pt]{IEEEtran}

\def\bb0{{\mathbb{0}}}

\def\bb{{\boldsymbol{b}}}

\def\b0{{\boldsymbol{0}}}

\def\b{{\mathrm{b}}}

\def\r0{{\mathbf{0}}}

\def\bsf0{{\bm{\mathsf{0}}}}

\def\N0{{N_{\mathrm{0}}}}

\def\bsf{{\boldsymbol{s}_\mathrm{f}}}

\newcommand{\be}{\begin{equation}}
\newcommand{\ee}{\end{equation}}
\newcommand{\bal}{\begin{align}}
\newcommand{\eal}{\end{align}}

\def\SNR    {{\mathsf{SNR}}}

\usepackage{graphicx}
\usepackage{amsmath}
\usepackage{subfigure}
\usepackage{mathtools}
\usepackage{epsfig}
\usepackage{float}
\usepackage{lipsum}
\usepackage{stfloats}
\usepackage{array}
\usepackage{amssymb}
\usepackage{cite}
\usepackage{url}
\usepackage{algorithm2e}
\usepackage{algorithmic}
\usepackage{multirow}
\usepackage{fancyh}
\usepackage[keeplastbox]{flushend}
\usepackage{upgreek}
\usepackage{dsfont}
\usepackage{gensymb}
\usepackage{setspace}

\usepackage{multicol, blindtext}
\usepackage{mathtools, cuted}

\usepackage{color}

\normalsize
\begin{document}

\title{\huge   Terahertz Line-Of-Sight MIMO Communication: \\Theory and Practical Challenges }
\author{Heedong Do, Sungmin Cho,  Jeonghun Park, Ho-Jin Song, \\ Namyoon Lee, and Angel Lozano \vspace{-3mm}
\thanks{H. Do, S.-M. Cho, H.-J. Song, and N. Lee are with
POSTECH, Pohang, Korea (e-mail:
\{doheedong, sungmin.cho, hojin.song, nylee\}@postech.ac.kr).  
J. Park is with Kyungpook Nat'l University, Daegu, South Korea (jeonghun.park@knu.ac.kr).  A. Lozano is with Univ. Pompeu
Fabra, 08018 Barcelona (e-mail: angel.lozano@upf.edu).
} }
\markboth{IEEE Communications Magazine, Draft}{}
\maketitle

\begin{abstract}
A relentless trend in wireless communications is the hunger for bandwidth, and fresh bandwidth is only to be found at ever higher frequencies. While 5G systems are seizing the mmWave band, the attention of researchers is shifting already to the terahertz range. In that distant land of tiny wavelengths, antenna arrays can serve for more than power-enhancing beamforming. Defying lower-frequency wisdom, spatial multiplexing becomes feasible even in line-of-sight conditions. This paper reviews the underpinnings of this phenomenon, and it surveys recent results on the ensuing information-theoretic capacity.  Reconfigurable array architectures are put forth that can closely approach such capacity, practical challenges are discussed, and supporting experimental evidence is presented.
\end{abstract}



%
\IEEEpeerreviewmaketitle

 \section{Introduction}
 
The two mechanisms whereby wireless systems can increase their bit rates are augmenting the bandwidth and improving the spectral efficiency, and both have progressed in tandem over the years. From 1G to 5G, the spectrum devoted to wireless communication has surged from a handful of MHz to multiple GHz, roughly three orders of magnitude, while system spectral efficiencies have risen by about two orders of magnitude. For 5G, microwave spectrum allocations no longer sufficed, and a first step is being taken beyond, into the mmWave realm (6--95 GHz). The spectral efficiency also continues to advance, despite the exhaustion of many of its classical improvement strategies, by virtue of multiple-input multiple-output (MIMO) massification \cite{boccardi2014five}. All in all, bit rates of about 20 Gb/s are within reach in 5G. Further headway towards the Tb/s milestone will again require a leap forward in both spectrum and spectral efficiency.

Spectrum-wise, the next frontier is the terahertz (THz) band, broadly defined as 0.1--10 THz and sandwiched between the mmWave and the far-infrared ranges. Although there are reasons why this band is largely unexplored, the obstacles look increasingly surmountable. First, signal sources of adequate power and room-temperature detectors of acceptable sensitivity have long been a major challenge, but there is promising progress on these fronts and recent experimental demonstrations with state-of-the-art solid-state electronics have reached 100 Gb/s over 20 GHz of bandwidth at 300 GHz \cite{THz120Gbps}. Second, the THz band is challenging in terms of radio propagation. In particular, and owing to the lack of diffraction, propagation is predominantly line-of-sight (LOS). This rules out wide-area coverage, yet it befits many emerging applications, both short-range in nature (inter-chip communication, datacenter interconnections, indoor local-area networks, kiosk downloads) and also of longer range (wireless backhaul, unmanned aerial vehicle (UAV) communication, satellite interconnection).  There are peaks of high atmospheric absorption that should be avoided, but, interspersed with those, there are enormous windows---hundreds of GHz altogether---where the absorption is below 10 dB/km \cite{akyildiz2014terahertz}.
 
Despite this vast amount of potential spectrum, the bandwidth that could be made available to individual users is curbed because:
\begin{itemize}
\item Power amplifiers are curtailed to about 10\% of the carrier frequency. 
\item The energy efficiency of analog-to-digital converters (ADCs) drops abruptly as the sampling rate pushes past 100 MHz \cite{murmann2016adc}.
While, up to that point, the ADC power consumption grows linearly with the bandwidth,  when moving from 100 MHz to 20 GHz the power consumption does not grow 200-fold but rather 10000-fold.  
\end{itemize}
It follows from these limitations on the per-user bandwidth that, as anticipated, the spectral efficiency remains important---yet there is
substantial downward pressure on it:
\begin{itemize}
\item The power of a signal source declines with increasing frequency and, above 300 GHz, it is capped below 20 dBm; this restricts the signal-to-noise ratio (SNR) and, as a result, the spectral efficiency.
\item The pathloss for omnidirectional antennas increases with frequency, further penalizing the SNR and the spectral efficiency.
\item The noise power grows with the bandwidth, compounding the SNR reduction.
\item To harness the ADC power expenditure, a sacrifice in resolution  may be  inevitable; with a lower resolution comes, again, a lower spectral efficiency.
\end{itemize}
Without a forceful countering of these effects, user bit rates may fall very short of the Tb/s mark. Besides, as in mmWave communication, the antidote to a low per-antenna spectral efficiency is to increase the number of antennas, capitalizing on the tiny wavelength to pack massive arrays onto compact form factors.  The most immediate use for such antenna arrays is SNR-enhancing beamforming with a single baseband chain. This is the solution currently adopted for mmWave.  While this is a sound starting point for THz communication as well, per-antenna baseband processing can already be envisioned.
 
The eventual conjunction of THz frequencies and per-antenna processing naturally invites MIMO transmission, which, while subsuming beamforming as a special case, broadens the scope to spatial multiplexing \cite{Foundations:18}. Moreover, as we move up in frequency, LOS goes from hampering spatial multiplexing to being an instrument for it.  This phenomenon is the thrust of the present paper, which expounds its theoretical foundations, discusses the chief practical obstacles, presents numerical and experimental evidence, and points to subsequent research.
 
\begin{figure*}
  \centering
  \includegraphics[width=1\linewidth]{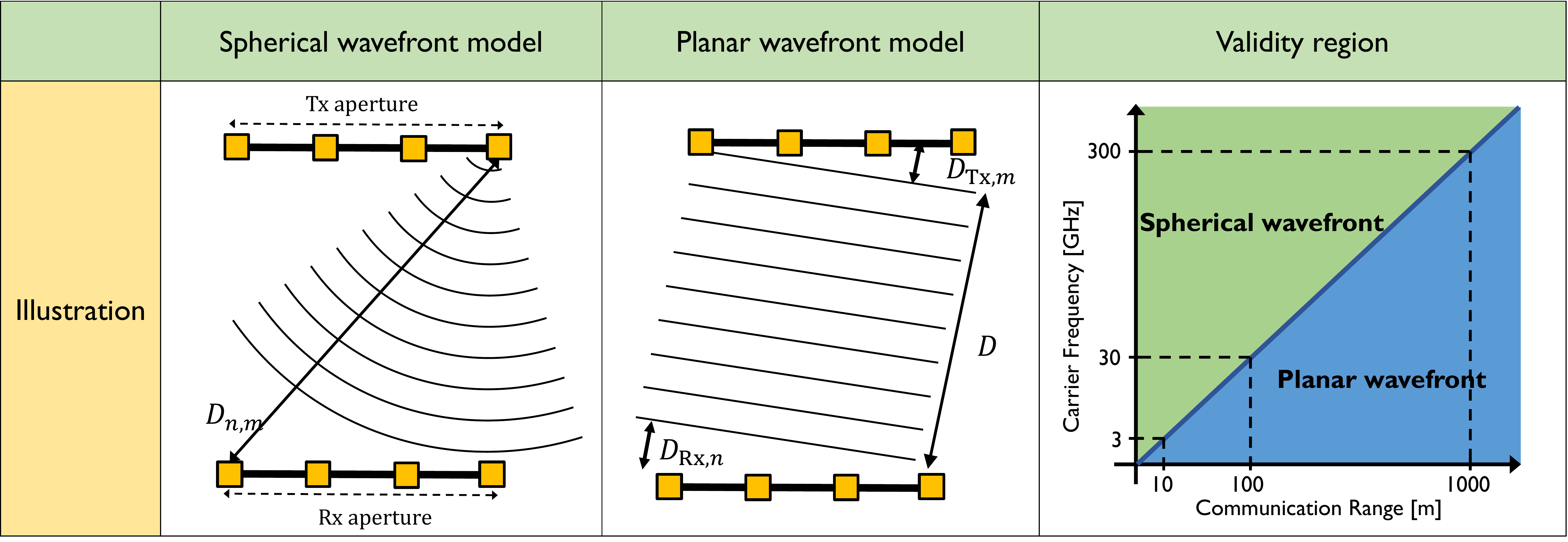}
  \caption{Spherical and planar wavefront illustrations, and respective validity regions as a function of the communication range and carrier frequency when the aperture of the transmit and receive arrays is 50 cm.}
  \label{Fig3}
\end{figure*} 
 
\section{THz LOS MIMO Communication}
\label{BernatCarla}

Commencing on the theoretical front, this section addresses three key aspects:
\begin{itemize}
\item The appropriate modeling of LOS MIMO channels at THz frequencies.
 \item Insights gleaned from applying information theory to such models. 
 \item Reconfigurable antenna architectures motivated by these insights. 
\end{itemize}

\subsection{Rethinking MIMO Channel Modeling for THz Frequencies}

Unlike at microwave frequencies, multipath propagation is very weak in the THz band because, compounding the lack of diffraction, the roughness of most surfaces is comparable to the wavelength. This leads to major scattering and reflection losses, leaving the LOS path as the dominant means of propagation.  Since microwave-frequency wisdom deems multipath richness as instrumental for spatial multiplexing while regarding LOS propagation as a roadblock to it, the possibility of spatial multiplexing would seem compromised at THz frequencies. Understanding why  this is not the case requires some careful modeling of the inherent channel features.

The magnitude and phase of the channel connecting the $m$th transmit with the $n$th receive antenna is governed by the corresponding distance, $D_{n,m}$. Under the reasonable proviso that the array apertures are small relative to their separation, all these distances are similar enough for the magnitudes of those responses to be taken as identical. In contrast, the phase of the channel responses cannot be deemed uniform,  as even minute distance variations may elicit pronounced phase differences. Hence, an LOS channel representation adopts the form of a matrix containing the phase of the responses for every transmit-receive antenna pair.

The time-honored model for this matrix representation regards the wavefronts as locally flat over each array. Geometrically, this corresponds to factoring out a common distance $D$ for all antenna pairs, as shown in Fig. \ref{Fig3}; what remains are then the phase shifts corresponding to the residual distances from each antenna to the wavefront, and, as evidenced in the figure, these are linear.  Mathematically, the planar wavefront approximation amounts to a first-order series expansion of every $D_{n,m}$ around $D$. As it turns out, the planar approximation is highly precise provided that the product of the transmit and receive array apertures is smaller than $4 \, \lambda D$, where $\lambda$ is the wavelength \cite{jiang2005spherical}.
Under this condition, which virtually always holds at lower frequencies, the LOS channel matrix is simply the outer product of the vectors of linear phase shifts induced by the plane wavefronts across the transmit and receive arrays. And, being the outer product of two vectors, the matrix is unit-rank, meaning that no spatial multiplexing (which requires a high-rank channel) is possible, but only beamforming (for which a unit-rank channel suffices).

While perfectly appropriate for microwave frequencies, the planar wavefront approximation becomes inadequate in THz territory. Because the wavelength is minute and the transmission distance tends to be short, the wave curvature over the arrays ceases to be negligible. Rather, such curvature creates a richer pattern of phase variations that endows the LOS channel matrix with a high rank.  Representing this reality requires the matrix  of absolute phases associated with the actual distances, $D_{n,m}$ (see Fig.~\ref{Fig3}).  While already noticeable at mmWave frequencies \cite{Bohagen:05,Sheldon:081, Bao:15, Halsig:172}, the effect of the wave curvature over the arrays becomes categorical on the THz band. This offers the opportunity of customizing the channel matrix through a careful arrangement of the antennas within each array. Indeed, rather than by the vagaries of multipath propagation, here the channel matrix is determined by sheer geometry.

Before further exploring the possibilities associated with custom antenna arrangements, it is instructive to view the problem at hand from the fundamental prism of information theory.
 
\subsection{Some Insights from Information Theory}

For every channel and SNR, the information-theoretic capacity is the highest possible spectral efficiency that can be achieved reliably. To incorporate the newfound ability of modifying the channel matrix through antenna position adjustments, we can introduce a broader capacity taken over all possible antenna arrangements at a given SNR. Characterizing this broader capacity entails identifying the best possible such arrangements at every SNR, a difficult task in general.

Fortunately, through a careful process of bounding it can be determined that the key role of the array geometries is to adjust the rank of the channel matrix depending on the SNR \cite{ISIT,do2020reconfigurable}. More precisely, the antenna dispositions should be such that a certain number of the singular values---the gains along the matrix preferred directions---are positive and identical while the rest are zero; the rank is the number of positive singular values. The design objective is thus to \emph{polarize} the singular values of the channel matrix into two states, positive and zero, depending on the SNR.
 
Three distinct regimes emerge:
 \begin{itemize}
\item At low SNR, when the communication is power-limited, the winning strategy is to construct the channel matrix enabling the highest possible power gain. There should be a single positive and large singular value, and beamforming should take place along the corresponding direction.
  \item At intermediate SNRs, a transition. The number of positive singular values should progress from one to the minimum number of antennas in the transmit/receive arrays as the SNR advances, and this number dictates the order of the spatial multiplexing. Indeed, as the SNR strengthens, multiple parallel signals can be sustained, multiplying the spectral efficiency.
 \item At high SNR, the singular values should all be positive and identical, and full spatial multiplexing of as many signals as antennas should take place.
 \end{itemize}

Besides being information-theoretically optimum, the equality of the positive singular values greatly simplifies the transmit and receive signal processing \cite{do2020reconfigurable}. Much of it amounts to applying unitary rotations, which are very amenable to implementation, and the allocation of power across the parallel transmit signals is trivially uniform.
 
 \begin{figure*}
  \centering
  \includegraphics[width=1.0\linewidth]{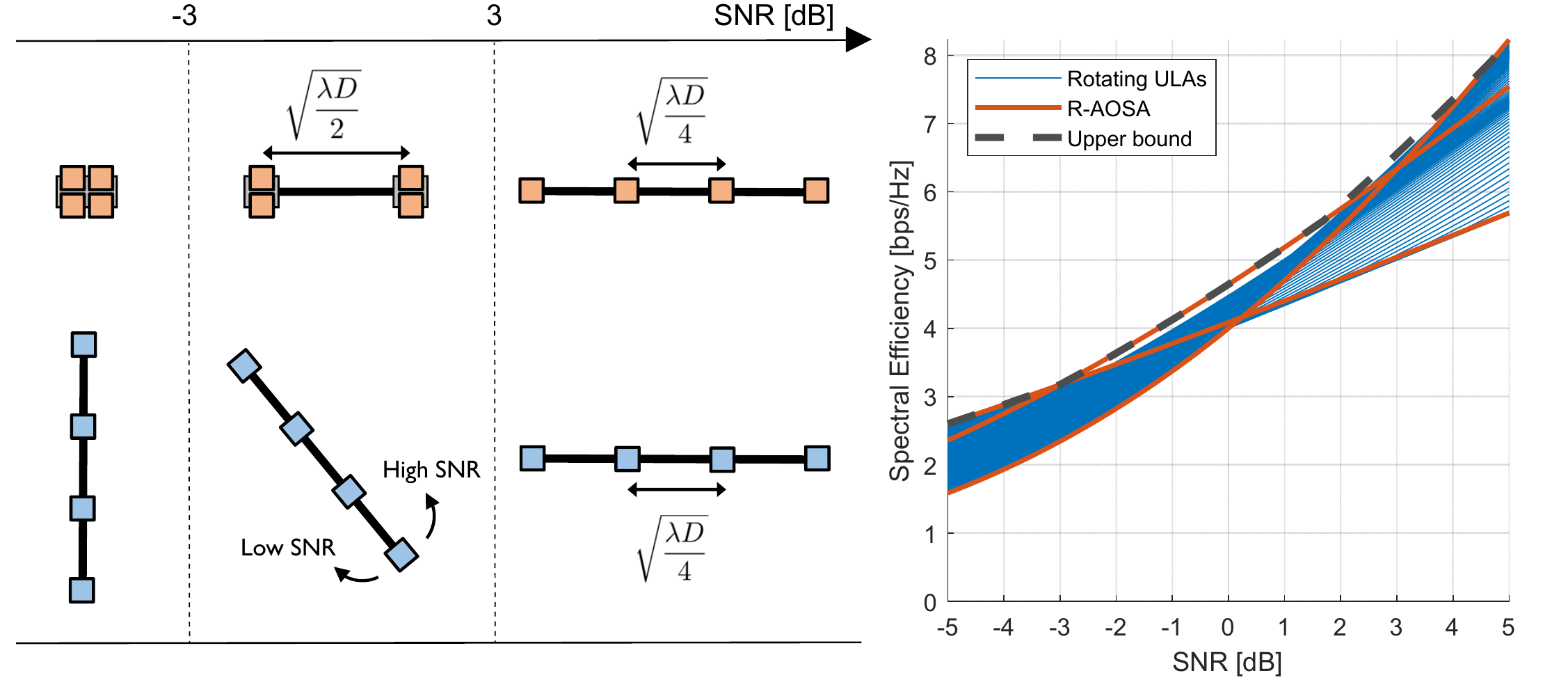}
   \caption{Left-hand side: R-AOSA and rotating ULA with four antennas as a function of the SNR; right-hand side: spectral efficiencies achievable by R-AOSAs and by rotating ULAs, respectively, versus a capacity upper bound.}
  \label{Fig4}
\end{figure*}

\subsection{Reconfigurable Array Architectures}

Having established the key design guideline for the arrays, let us now put forth two specific architectures that, inspired by this guideline, polarize the singular values as a function of the SNR. These are the reconfigurable array-of-subarrays (R-AOSA) and the rotating ULA.

\subsubsection{R-AOSA}

An AOSA consists of $r$ subarrays, uniformly spaced. The antennas are divided evenly among these subarrays at, within each, they are tightly spaced relative to the wavelength. With AOSAs at both ends, the channel matrix features $r$ positive and equal singular values. By progressively increasing $r$ as the SNR improves, i.e., by having more subarrays with fewer antennas each, an AOSA is rendered reconfigurable. At the low-SNR extreme, a linear R-AOSA reverts to a single array of tightly spaced antennas whereas, at the high-SNR end, it becomes a ULA with the precise inter-antenna spacing that makes all the singular values equal \cite{Driessen:99}. This is illustrated, for a four-antenna linear R-AOSA, in Fig. \ref{Fig4}. Up to an SNR of about $-3$ dB, the AOSA adopts the form of a single compact array, subsequently morphing to twin two-antenna arrays, and, beyond $3$ dB, to a four-antenna ULA. The spectral efficiencies achievable by these three configurations are also depicted in Fig. \ref{Fig4}, alongside a capacity upper bound. By switching to the correct configuration at every SNR, the bound is tracked rather closely.  This excellent performance, however, does come at the cost of having to physically rearrange the antennas whenever the SNR changes, a drawback addressed by the next architecture.

\subsubsection{Rotating ULA}

 As it turns out, the singular values of the channel matrix can also be manipulated by simply rotating the transmit and/or receive arrays \cite{do2020reconfigurable}. Moreover, provided the number of antennas is minimally large, an adequate rotation can essentially polarize the singular values in any desired fashion. (Asymptotically in the numbers of antennas, the polarization is exact; for finite numbers thereof, it is approximate.) This is again illustrated, for a four-antenna ULA, in Fig \ref{Fig4}. For SNRs below -$3$~dB, the ULA should be aligned with the direction of transmission, subsequently spinning towards a broadside disposition, reached at about $3$~dB.The exact rotation depends, not only of the SNR, but also on the number of antennas. The performance achievable on a set of angles is exemplified in Fig. \ref{Fig4} alongside the capacity upper bound; by effecting the correct rotation at every SNR, the bound is virtually attained.
 
Most interestingly, the rotating ULA architecture need not be embodied by a mechanically turnable array. Instead, it can also be realized by electronically selecting among various ULAs having a radial disposition. While additional antennas are needed to implement this solution, the number of radio-frequency chains stays the same, meaning that the additional cost is marginal. And, as it turns out, with as few as three fixed ULAs, properly angled, the performance is hardly different from that of a freely rotating ULA \cite{do2020reconfigurable}.

\begin{figure}
\centering
\includegraphics[width=1\linewidth]{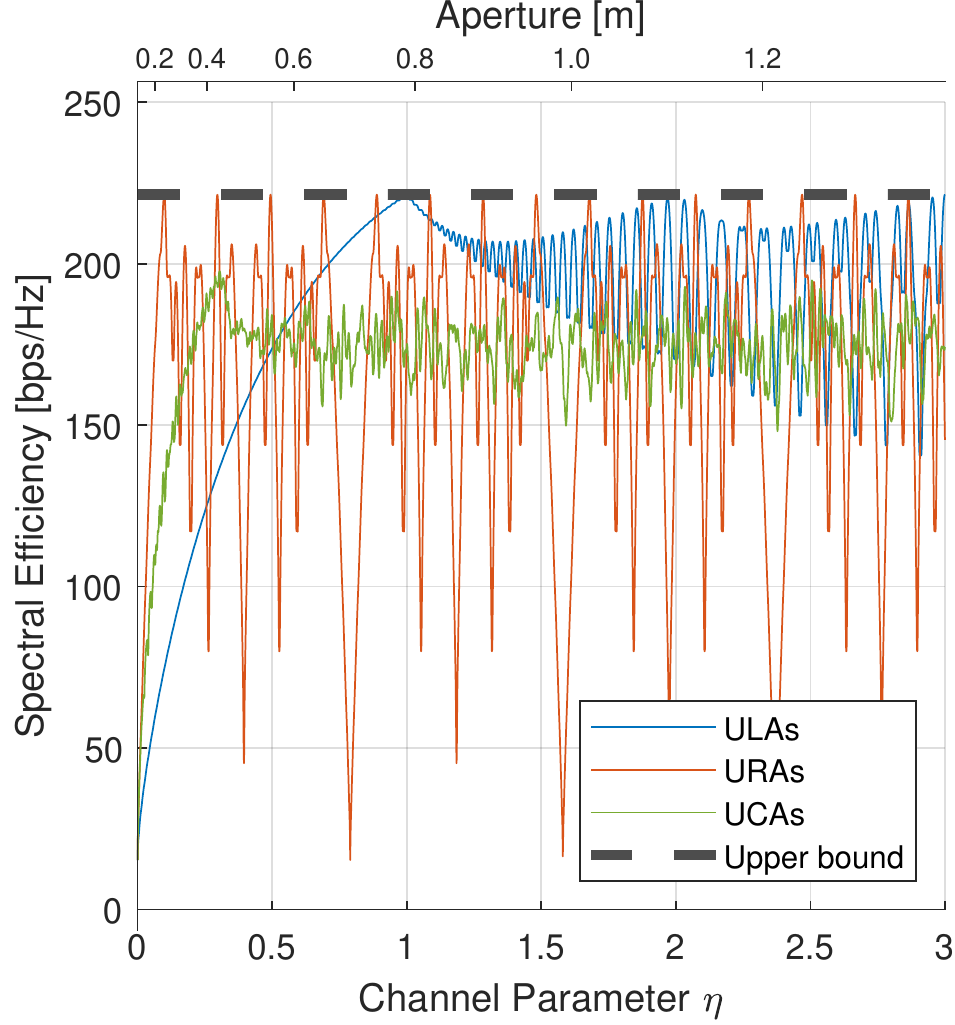}
 \caption{ 
Spectral efficiency as a function of the channel parameter for ULAs, URAs, and UCAs, with $64$ transmit and receive antennas at $\SNR=10$ dB.  Also shown is the capacity upper bound at this SNR. The array apertures---equal at transmitter and receiver---corresponding to the channel parameter are also indicated for a transmission distance of 10 m. }
\label{Fig5}
 \end{figure}

\section{Beyond ULAs: Rectangular \\ and Circular Arrays}

While we have been invoking ULAs (and AOSAs, which can be interpreted as two-tier ULAs) as a running example, the ideas discussed hitherto extend to any other array type. In particular, a popular alternative to the ULA is the uniform rectangular array (URA), which in essence is the cartesian product of two ULAs, and the uniform circular array (UCA), which is the preferred geometry for a sister communication technique based on orbital angular momentum  \cite{OAM2012}. Both the URA and the UCA feature more compact form factors than the ULA for a given number of antennas. To facilitate a comparison of these and other array types in LOS environments, it is convenient to define the channel parameter $\eta =  \text{(Tx aperture)} \text{(Rx aperture)} / ( \lambda D N_{\sf min} ) $,
which blends the array apertures, the wavelength, the transmission distance, and the minimum number of antennas in the transmit/receive arrays, $N_{\sf min}$, into a single quantity; as it happens, the LOS MIMO capacity depends only on this quantity \cite{do2020reconfigurable}. For a ULA, the array aperture in the above relationship is the length, for a URA it is the side of the square, and for a UCA it is the diameter of the circle on which the antennas are placed;
in all cases, the aperture is measured with the arrays broadside to the transmission.

Equipped with the aforedefined channel parameter, we can contrast the spectral efficiencies of ULA, URA, and UCA architectures over a broad assortment of configurations. A comparison is presented in Fig. \ref{Fig5} for $64$ antennas at each end of the link and $\SNR=10$ dB, with the channel parameter sweeping the interval $[0,3]$. This corresponds to varying either the transmission distance, the wavelength, the number of antennas, or the apertures (say by reconfiguring the arrays, if they are embodied by R-AOSAS, or by rotating them, thereby modifying their broadside projection).
With ULAs, the capacity upper bound is essentially achieved for specific parameter values.  Applying the ideas in the previous section, any ULA can be reconfigured to operate at these points through a simple rotation---with the smallest such operating point being the most attractive from the vantage of a small array footprint. URAs also attain the upper bound for specific configurations, although in this case the reconfigurability need not correspond to a mere rotation. As of UCAs, there is no configuration for which they do not exhibit a certain shortfall with respect to capacity, suggesting that the circular geometry is less favorable for the purpose of polarizing the singular values. And, as a fallout of an incomplete polarization, the optimum allocation of powers across the parallel transmit signals is not uniform.
In exchange, UCAs do simplify some of the processing: rather than channel-dependent unitary rotations, here the transmitter and receiver reduce to channel-independent Fourier matrices.

\section{Practical Challenges }

The implementation of THz MIMO transmission is not without challenges, some of which are sketched next.

\subsubsection{Misalignments}
Maintaining a perfect alignment between transmitter and receiver may be impossible in some of the envisioned applications. Analyses of the sensitivity to  unwanted tilts and rotations are necessary to guide the designs and to establish the accuracy to which the ideal orientations must be tracked in the face of motion or disturbances, say wind turbulence in the case of UAVs.
 
 \subsubsection{Frequency Variations}
  The optimum configuration at some wavelength need no longer be optimum at some other wavelength, everything else being the same. A meaningful objective is thus to identify arrangements that are robust across the broadest possible swaths of spectrum.

\subsubsection{Intersymbol Interference}
In stark contrast with lower frequencies, where LOS channels are devoid of intersymbol interference, at THz frequencies it may arise on account of the short transmission distance. Precisely, back-and-forth reflections may occur when the arrays face each other at close range, leading to a rather distinct form of distance-dependent intersymbol interference. This issue needs to be characterized, and mitigating solutions developed.

\subsubsection{Low-Resolution ADCs}
Lowering the resolution of the ADCs and of the mirror digital-to-analog converters ameliorates the cost and the power consumption, but at the expense of a surge in quantization noise. While some results are available on the capacity of MIMO channels with one-bit ADCs the understanding is far from complete \cite{nam2019capacity}. Substantial progress might be possible for LOS MIMO specifically, given its rather unique nature.

\subsubsection{Array Apertures}
 
Having reduced array footprints is a desirable trait for certain applications, say portable devices or UAVs. At a given SNR, the optimum channel parameter maps to a product of the transmit and receive apertures that grows with the transmission distance, meaning that long-range transmissions are in principle problematic from the standpoint of keeping the apertures small. However, there is relief in that the dependence is on the product of the apertures, making it possible to keep one end of the link compact at the expense of the other end. Depending on the setting, then, the correct balance of transmit and receive apertures would have to be determined.

\begin{figure*}[htp]
  \centering
  \includegraphics[width=1\linewidth]{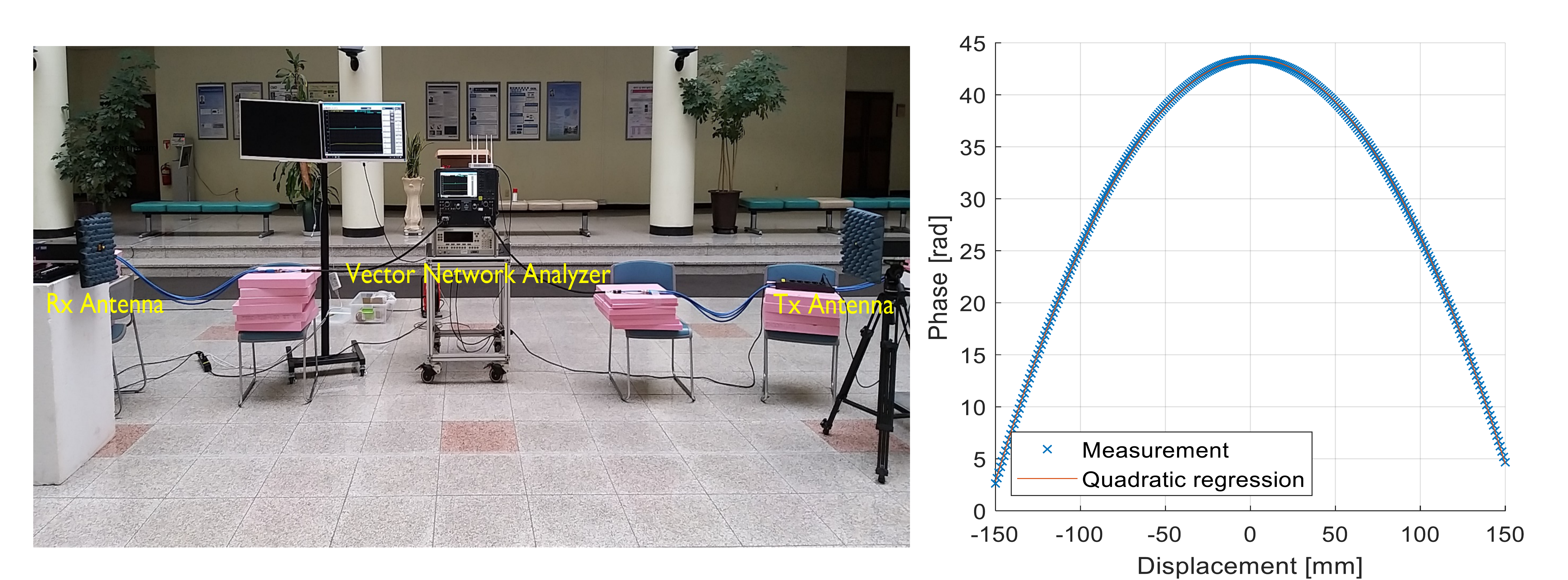}
  \caption{Left-hand side: experimental setup; right-hand side: phase vs receive antenna displacement.
}
  \label{Fig6}
\end{figure*}

\section{Experimental Evidence}

Shown in Fig. \ref{Fig6} is an experimental setup designed to verify some of the underlying ideas. A narrowband signal is transmitted over a link distance $D$ ranging from 0.5 m to 5.4 m at 300 GHz.  Both transmitter and receiver are equipped with a single horn antenna, fairly directive (25 dBi antenna gain for 18 degrees of half-power beamwidth). This directivity ensures that no significant ground bounce is present while radiation-absorbent material prevents back-and-forth reflections. The measurements rest on a through-reflect-line calibration kit and a vector network analyzer, with a minimum SNR of $30$~dB at all times.

Through a mechanical displacement of the receive antenna in 1-mm steps, with a dwell time of 5 ms, a 300-antenna synthetic ULA receiver is produced while the transmitter is held fixed. Despite the directivity of the physical antennas, it is verified that the magnitude of the channel response varies but little over the synthetic array; in most deployments, the antennas would be far less directive, reinforcing this aspect. With the premise that the channel behavior depends only on the phase fluctuations being therefore satisfied, the phase of the channel response at each displacement, i.e., for each virtual receive antenna when $D=1.8$ m, is depicted in Fig. \ref{Fig6}. It is seen that the phase exhibits a quadratic variation over the synthetic ULA, in validation of the spherical wavefront model; under a planar wavefront model, instead, it would be linear.

\section{Summary}
 
The vast amount of spectrum available on the THz band dwarfs what is currently being conquered at mmWave frequencies---even with all the intervals of high atmospheric attenuation discounted. However, this enormous bandwidth can only be assigned in moderate doses to individual users, and thus the spectral efficiency remains a relevant figure of merit vis-a-vis bit rates. As an assortment of factors pressure the spectral efficiency down at these frequencies, a mechanism to boost it becomes imperative, and MIMO emerges as a welcome opportunity.

Contrary to the lower-frequency wisdom that only beamforming is possible in the LOS conditions that are prevalent in THz channels, the combination of a tiny wavelength and a short transmission distance opens the door to spatially resolving each antenna, and hence to spatial multiplexing, even in LOS situations.Transcending the classic planar wavefront approximation and embracing the actual spherical nature of the wave propagation, it is possible to tightly bound the information-theoretic capacity of LOS MIMO settings and observe that the optimum transmission strategy polarizes the channel's singular values depending on the SNR. Several ways to go about this have been put forth, chiefly the SNR-dependent rotation (either mechanic or electronic) of a ULA.

The rollout of MIMO transmission at THz frequencies is not without challenges, and the main practical hurdles have been outlined. Finally, experimental evidence has been provided to support some of the theoretical underpinnings.


\section*{Acknowledgments}

This work was supported by the Samsung Research Funding and Incubation Center of Samsung Electronics under Project SRFC-IT1702-04, by the Basic Science Research Programs under the National Research Foundation of Korea (NRF) through the Ministry of Science and ICT under NRF2020R1C1C1013381, and by the European Research Council under the H2020/ERC grant agreement 694974.

  
  


  \bibliographystyle{IEEEtran}

    \bibliography{main_AL_5.bib}

\vspace{1cm}

\begin{IEEEbiographynophoto} 
	{Heedong Do} (S'20) received a B.S. in mathematics and a M.S. in electrical engineering from Pohang University of Science \& Technology (POSTECH), Pohang, Korea, in 2018 and 2020, respectively. He is currently a Ph.D. candidate.
\end{IEEEbiographynophoto}

\vspace{0.1cm}

\begin{IEEEbiographynophoto}
 {Sungmin Cho} (S'19) received a B.S. from Kyungpook Nat'l University, Daegu, Korea in 2017 and a M.S degree in 2019 from Pohang University of Science \& Technology (POSTECH), where he is currently pursuing the Ph.D. degree.
\end{IEEEbiographynophoto}
 
 \vspace{-0.1cm}

 \begin{IEEEbiographynophoto}
 {Jeonghun Park} (S'13--M'17) is an Assistant Professor at Kyungpook Nat'l University, Daegu, Korea. Before that, he worked at Qualcomm wireless R\&D, San Diego, USA. He received a Ph.D. degree in electrical and computer engineering at The University of Texas at Austin, USA, in 2017.
\end{IEEEbiographynophoto}

\vspace{-0.1cm}

\begin{IEEEbiographynophoto}
 {Ho-Jin Song} (S'02-M'06-SM'13) received a Ph.D. degree from Gwangju Institute of Science \& Technology (GIST), Gwangju, Korea, 2005. He was with the Nippon Telegraph and Telephone (NTT) Laboratories, Kanagawa, Japan, in 2006 - 2016. Since 2016, he has been an Associate Professor at POSTECH, Pohang, Gyeongbuk, Korea.  Prof. Song was awarded the Best Thesis Award from the Gwangju Institute of Science and Technology (2005), the Young Scientist Award of the Spectroscopical Society of Japan (2010), the IEEE MTT-S Tatsuo Itoh Prize (2014), and the Best Industrial Paper Award at IEEE MTT-S-IMS2016 (2016).
\end{IEEEbiographynophoto}

\vspace{-0.1cm}

 \begin{IEEEbiographynophoto}
{Namyoon Lee} (S'11--M'14--SM'20) received a Ph.D. degree from The University of Texas at Austin, in 2014.  He was with Communications and Network Research Group,  Samsung Advanced Institute of Technology (SAIT), Korea, in 2008-2011 and also with Wireless Communications Research (WCR), Intel Labs, Santa Clara, USA, in 2015-2016. He is currently an Associate Professor at POSTECH, Pohang, Gyeongbuk, Korea. He was a recipient of the 2016 IEEE ComSoc Asia--Pacific Outstanding Young Researcher Award. He is currently an Editor for both the IEEE Trans. on Wireless Communications and the IEEE Communications Letters.
\end{IEEEbiographynophoto}


\begin{IEEEbiographynophoto}
{Angel Lozano}(S'90 -- M'99-- SM'01 - F'14) received a Ph.D. from Stanford University in 1998. From 1999 to 2008 he was with joined Bell Labs (Lucent Technologies, now Nokia). He is currently a Professor at Univ. Pompeu Fabra (UPF), Barcelona, and the co-author of the textbook ``Foundations of MIMO Communication'' (Cambridge University Press, 2019). He serves as Area Editor for the IEEE Trans. on Wireless Communications. He received the 2009 Stephen O. Rice Prize, the 2016 Fred W. Ellersick prize, and the 2016 Communications Society \& Information Theory Society joint paper award. He holds an Advanced Grant from the European Research Council and was a 2017 Clarivate Analytics Highly Cited Researcher.

\end{IEEEbiographynophoto}

\end{document}